# Radio Communication Scenarios in 5G-Railways

Ruisi He[1], *Senior Member, IEEE*, Bo Ai*[1], *Senior Member, IEEE*, Zhangdui Zhong[1], *Senior Member, IEEE*, Mi Yang[1], *Student Member, IEEE*, Chen Huang[1], *Student Member, IEEE*, Ruifeng Chen[2], Jianwen Ding[1], Hang Mi[1], Zhangfeng Ma[1], *Student Member, IEEE*, Guiqi Sun[1], and Changzhu Liu[1]

[1]State Key Lab of Rail Traffic Control and Safety, Beijing Jiaotong University, Beijing, China

[2]Institute of Computing Technology, China Academy of Railway Sciences, Beijing, China

*Abstract*—With the rapid development of railways, especially high-speed railways, there is an increasingly urgent demand for new wireless communication system for railways. Taking the mature 5G technology as an opportunity, 5G-railways (5G-R) have been widely regarded as a solution to meet the diversified demands of railway wireless communications. For the design, deployment and improvement of 5G-R networks, radio communication scenario classification plays an important role, affecting channel modeling and system performance evaluation. In this paper, a standardized radio communication scenario classification, including 18 scenarios, is proposed for 5G-R. This paper analyzes the differences of 5G-R scenarios compared with the traditional cellular networks and GSM-railways, according to 5G-R requirements and the unique physical environment and propagation characteristics. The proposed standardized scenario classification helps deepen the research of 5G-R and promote the development and application of the existing advanced technologies in railways.

## I. INTRODUCTION

DURING the past decades, global demand for wireless communication networks has been rapidly increasing, mainly due to the continuing growing numbers of wireless users and new types of services and devices. The future communication system is supposed to provide seamless and ubiquitous on-demand coverage for various application scenarios [1]-[3]. Communication scenario is one of the prior concerns for communication system design, network implementation, and optimization of quality of services (QoSs). A proper communication scenario determination significantly benefits radio propagation characterization, network design and deployment, resource allocation and optimization, and improve system performance and robustness in different communication environments.

Considering that different communication environments have different physical radio propagation features and different requirements of communication services, many standard organizations, such as the 3rd Generation Partnership Project (3GPP) and International Telecommunication Union (ITU), make the decriminalization of communication scenarios the primary work when developing the next generation communication system. Generally, there are several universal principles to determine communication scenarios:

*(1) Communication Service Requirements*

The main goal of communication scenario classification is to better improve the communication services by optimizing system design, network coverage, network capacity, and communication latency. Meanwhile, different types of communication services may be provided in different scenarios, and they usually have different key performance indicators, e.g, collision avoidance system in vehicle-to-vehicle is sensitive to network latency, whereas file transfer protocol services have high requirement for network capacity [4]. Hence, it is necessary to determine communication scenario based on complete consideration of communications service requirements. Following the idea, 3GPP classifies communication scenarios into five categories in TR 22.891 [5]: enhanced mobile broadband, critical communications, massive machine-type communications, network operation, and enhancement of vehicle-to-everything, which are further defined into 74 specified scenarios according to the requirements of different services. The description, potential service requirements, and potential operational requirements of each case are given in detail. A comprehensive scenario separation benefits the network resource allocation and optimization, however, it also brings redundant and exhausting work of network design and channel characterization. Therefore, in 2020, 3GPP has simplified and updated the list of communication scenarios. By considering the ITU-Radiocommunica (ITU-R) discussion on International Mobile Telecom System-2020 (IMT-2020) requirements, 3GPP identifies the typical deployment scenarios associated with attributes such as carrier frequency, inter-site distance, user density, maximum mobility speed, etc, and uses the scenarios to further develop requirements for the next-generation access technologies. In TR 38.913, the communication scenarios are determined by considering different principles [6]: (i) indoor hotspot, dense urban, rural, urban macro, and highway scenario are determined according to physical communication environment; (ii) high speed, extreme long-distance coverage in low-density areas, urban grid for the connected car, commercial air to ground scenarios, light aircraft scenario, and satellite extension to terrestrial are determined based on the types of communication devices and services.

Except for 3GPP, ITU, as one of the biggest standard organizations, proposes the IMT-2000 [7] that defines a range of radio operating environments based on three attributes: network attributes, physical attributes, and user attributes, which lead to 9 terrestrial scenarios and 4 satellite scenarios including business indoor, neighborhood indoor/outdoor, home, urban vehicular, urban pedestrian outdoor, rural outdoor, terrestrial aeronautical, fixed outdoor, local high bit rate environments, urban satellite, rural satellite, satellite fixed-mounted, and indoor satellite environments. Similar to 3GPP, ITU proposes IMT-2020 [8] by simplifying the scenarios according to requirements of the new generation of communication services and Internet of things (IoT) network, which includes: (i) general wide coverage scenario, which requires a throughput rate of more than 100 Mbps even at the edge of the cell; (ii) high capacity hot spot, which requires a throughput rate of 1-10 Gbps and with a throughput density of more than 10 Tbps/km$^2$; (iii) low latency and high-reliability scenario, which faces to vehicle-to-vehicle communications and industrial network and provides a device-to-device communication latency lower than milliseconds with nearly 100% reliability; and iv) low-power consumption with massive connection scenario, which meets the requirements of typical sensor network and provides a low-power consumption, low cost and numerous connection with wide coverage.

Similarly, the Next Generation Mobile Networks (NGMN) [9] has developed twenty-five use cases for 5G, as representative communication scenarios, which are grouped into eight scenario series. The scenarios and scenario series serve to meet the stipulating requirements and define building blocks of 5G architecture. In this case, the defined specified scenarios are not meant to be exhaustive, but rather as a tool to ensure that the level of flexibility required in 5G is well captured.

*(2) Radio Propagation Characteristics*

Radio propagation characteristics depend on topographical and electromagnetic features of environment where communication system is operating. Generally, wireless channels show different radio propagation characteristics in different physical environments, due to the interactions between radio wave and objects during propagation of signal. A series of propagation scenarios need to be predefined before modeling the specific channels, despite the communication service categories. Following this idea, there are numerous standards and channel models define propagation scenarios according to physical feature of radio environment.

The WINNER project in Europe was established in 2004. Based on Universal Mobile Telecommunications System (UMTS) and IMT-2000 scenario definitions, the WINNER project defines four typical scenario series including indoor and outdoor office building, urban micro-cell and indoor hot spot, urban and suburban macro-cell, and rural scenarios. Seventeen specified scenarios are further determined considering the relative position between transmitter (Tx), receiver (Rx), and the physical environment for the four scenario series. Different from 3GPP and ITU who tend to simplify communication scenarios by considering service categories, the WINNER II project follows the original four scenario series and further specifies and details the cases into eighteen specified scenarios [10]. The scenarios cover most typical propagation cases. They are not intended to cover all possible environments and conditions, e.g., the mountainous or even hilly rural environments have not been covered. Similarly, the antenna heights do not cover all possible values that could be seen. Generally, the environments are such that are found in urban areas of European and North American countries.

On the other hand, instead of directly classifying scenarios based on physical environment, the COST 259 channel model defines communication scenarios into a hierarchical structure with three main layers [11]: cell types, radio environment, and propagation scenarios. The cell type is the first and essential distinction categories including macro-, micro-, and pico-cells, whereas the layer of radio environment highly depends on physical condition, e.g., urban, suburban, hilly terrain, or office. In addition, it also considers the quality of connection, e.g., bad urban macro/micro scenario. The bottom layer of the hierarchical structure consists of propagation scenario (PS), which is defined as random realization of multipath condition and highly depends on the geometry relationship between Tx/Rx and scatterers in the channels. Following the same structure, COST 2100 [12] extends COST 259 to cover MIMO systems, including multi-user, multi-cellular, and cooperative aspects, without requiring a fundamental shift in the original modeling philosophy.

Generally speaking, there are overlaps between the above two ideas of classifying communication scenarios. A specific environment, e.g., indoor office, usually has specific propagation channel characteristics (e.g., massive reflective and scattering objects) and specific type of communication service (e.g., hot spot network). However, despite different philosophies of defining scenarios, most of them are based on physical insight and relatively subjective communication engineering experience. 5G wireless network defines three main application scenarios: enhanced mobile broadband (eMBB), massive machine-type communication (mMTC), and ultra-reliable and low latency communication (uRLLC). Nowadays, 6G enhances the main application scenarios of 5G, and extends them to more diverse scenarios such as space-air-ground-sea integrated networks [1]. It is still a major challenge to specify detailed communication scenarios by taking into account the requirements of network services and the radio propagation environments.

## II. RADIO COMMUNICATION SCENARIOS IN GSM-R

GSM-R system has been widely used for railway communications, and its scenario feature and classification are important references for 5G-railways (5G-R). In the previous works, [13] and [14] propose radio propagation scenario partitioning for GSM-R system [15], which is based on a large number of railway coverage measurements carried out in China. Scenario classification in [13] and [14] is mainly considers physical propagation characteristics of environments, such as propagation environment structure, scatterer types, density of multipath components (MPCs), and distance of scatterer away from train, etc. The purpose is to distinguish the scenarios with different propagation mechanisms and channel characteristics, to facilitate propagation channel characterization, channel modeling, and communication system design. In [13], the typical railway scenarios for GSM-R are classified into twelve scenarios as follows.

*1. Viaduct Scenario*. Viaduct is one of the most common scenarios in railway communications, especially for high-speed railways (HSRs). Viaduct is a long bridge-like structure, and the height is mostly between 10-30 m. Generally, few scatterers can obstruct the communication link between base station (BS) and train, and an important feature of viaduct channel is line-of-sight (LOS) propagation. Only for a few cases, tall trees and buildings close to rail track may lead to none-LOS (NLOS) propagation in viaduct.

*2. Cutting Scenario*. Cutting often occurs in non-plain areas, which is used to make train pass through large obstacles (such as hills) smoothly on uneven ground. The U-shaped structure of cutting leads to rich reflections and scatterings from both sides of rail, most of which are concrete-fixed hillsides or naturally formed mountains. Mostly, LOS propagation exists in cutting scenario, however, the cross-bridge in cutting leads to NLOS propagation and the reduced communication quality.

*3. Tunnel Scenario*. Tunnel is an artificial closed environment and widely used in railways. The sectional view of tunnel in railways is usually vaulted or semicircle, with a height of 5-10 m and a width of 10-20 m. The length of tunnel mostly ranges from several to dozens of kilometers. Due to the smooth walls and the close structure of tunnel, there are rich reflection and scattering components inside tunnel, where wave guide effect dominates radio propagation. In addition, leaky coaxial cable is widely used for wireless coverage in tunnels, which is different from BS-to-train link in open scenarios.

*4. Station Scenario*. Railway station is an important railway facility and it generally consists of a platform close to rail track and a depot providing related services. Based on capacity of transportation, railway stations can be generally divided into three categories: medium- or small-sized station, large station, marshaling station and container depot. Railway station is like indoor propagation scenario. The BSs are mostly located outside awnings, and sometimes inside awnings. This special structure may lead to NLOS propagation and rich MPCs, and has significant impacts on propagation characteristics. In addition, the metallic carriages in marshaling station and container depot usually result in complex multipath structure and rich reflection and scattering components.

*5. Composite Scenario*. Composite scenario refers to the frequent switching of several different propagation scenarios in one communication cell. There are usually two typical composite scenarios: tunnel group and cutting group. When trains pass through mountainous areas, tunnel groups are widely present, i.e., train frequently moves in and moves out of tunnels. Frequent changes from tunnel to open area will greatly increase severity of channel fading and result in reduced communication quality. For cutting group, the steep walls on both sides may disappear temporarily, resulting in frequent scenario changes. This will make it challenging to predict wireless signal and affect quality of wireless communication.

*6. In-Carriage Scenario*. In-carriage scenario includes two categories: relay transmission and direct transmission. Relay transmission occurs where wireless coverage is provided by moving relay stations, and the propagation in carriage can be covered by some indoor channel models. Direct transmission refers to the situation of providing high-quality communication between BS and the users in carriages through wireless link. In this case, the carriage penetration loss has a significant influence on radio propagation.

*7. General Scenarios*. General scenarios refer to those scenarios widely used in the traditional cellular networks and also appear in railway environments, including urban, suburban, rural, water, and mountain, etc. Trains usually move in urban or suburban scenarios when passing through cities or towns. There usually exist buildings within 50-100 m around rail track. Rural scenario usually has many open areas. There are almost no building around the track and scatterers are rare in rural. River scenario refers to the situation where a train crosses a river or runs along one side of the river. Generally, there is no scatterer on water surface and obvious reflections will be caused by the surface. Mountain scenario widely occurs when there are mountains or hills around rail track, where BS tower is usually placed on the mountain. It should be noted that these general scenarios often form various composite scenarios with the special railway scenarios. For example, since stations are generally located in urban, urban-station composite is quite common. In addition, mountain scenario is often accompanied by cutting or viaduct.

Propagation scenario classification for GSM-R system plays an important role in analysis of radio propagation mechanism, wireless channel modeling, and performance evaluation of key railway communication technologies. It is also the foundation of BS deployment, wireless network planning and optimization. A reasonable and effective scenario classification is helpful to establish accurate path loss and shadow fading models for wireless link budget. If special railway scenarios are not well considered, there will be significant errors in link budget. A refined GSM-R scenario classification can be used to develop standardized channel model. The results in [14] show that by using the standardized channel model based on accurate scenario classification, the accuracy of link budget can be significantly improved.

III. RADIO COMMUNICATION SCENARIOS IN 5G-R

For the deployment and improvement of 5G-R networks, the radio propagation scenario classification plays an important role, and it significantly affects channel modeling and system performance evaluation [16]. In [13], propagation scenarios for the GSM-R system are defined, however, they cannot be directly applied to 5G-R as the applications and frequency bands are different. Therefore, it is necessary to establish a detailed propagation scenario classification for the 5G-R system. A few existing 5G-R scenarios classifications focus on application scenario classification. For example, in [17], the application scenarios are divided into mission critical communication and passenger communication. Among them, the mission critical communication includes train-to-ground communication and train communication network. The requirements for data rate, latency, reliability, availability, and maintainability for each application scenario are introduced, and detailed parameters are shown for these indicators. However, there is still limited investigation on propagation scenario classification of 5G-R. In this paper, by considering both propagation characteristics and new applications/services, a standardized radio communication scenario classification is proposed for 5G-R, including 18 scenarios. Table 1 and Table 2 summarize the key attributes and characteristics of each scenario. Details for each scenario are presented as follows.

TABLE 1
PROPOSED STANDARDIZED RADIO COMMUNICATION SCENARIO CLASSIFICATION AND THE CORRESPONDING KEY ATTRIBUTES.

| No. | Scenario | Experienced Data Rate | Connection Density | Reliability | Delay | Service | 5G-R Application |
|---|---|---|---|---|---|---|---|
| 1 | Viaduct | 0.1-1 Gbps | $10^3$-$10^4$/km | High | 10-100 ms | Control signal, voice, data, video, etc. | Intelligent operation |
| 2 | Cutting | 0.1-1 Gbps | $10^3$-$10^4$/km | High | 10-100 ms | Control signal, voice, data, video, etc. | Intelligent operation |
| 3 | Tunnel | 0.1-1 Gbps | $10^3$-$10^4$/km | High | 10-100 ms | Control signal, voice, data, video, etc. | Intelligent operation |
| 4 | Urban | 0.1-1 Gbps | $10^3$-$10^4$/km | High | 10-100 ms | Control signal, voice, data, video, etc. | Intelligent operation |
| 5 | Suburban | 0.1-1 Gbps | $10^3$-$10^4$/km | High | 10-100 ms | Control signal, voice, data, video, etc. | Intelligent operation |
| 6 | Rural | 0.1-1 Gbps | $10^3$-$10^4$/km | High | 10-100 ms | Control signal, voice, data, video, etc. | Intelligent operation |
| 7 | Mountain | 0.1-1 Gbps | $10^3$-$10^4$/km | High | 10-100 ms | Control signal, voice, data, video, etc. | Intelligent operation |
| 8 | Water | 0.1-1 Gbps | $10^3$-$10^4$/km | High | 10-100 ms | Control signal, voice, data, video, etc. | Intelligent operation |
| 9 | Platform | 0.1-10 Gbps | $10^4$-$10^5$/km$^2$ | High | 1-10 ms | Control signal, voice, data, video, etc. | Smart station |
| 10 | In-Carriage | 0.1-10 Gbps | $10^3$-$10^4$/km | Moderate | 1-10 ms | Voice, data, video, etc. | Intelligent equipment |
| 11 | Train-to-Train | 0.01-0.1 Gbps | - | High | 1-10 ms | Control signal | Intelligent operation, intelligent equipment |
| 12 | Station | 0.1-10 Gbps | $10^4$-$10^5$/km$^2$ | Moderate | 1-10 ms | Voice, data, video, etc. | Smart station |
| 13 | Railway Yard | 0.1-1 Gbps | $10^4$-$10^5$/km$^2$ | High | 1-10 ms | Control signal, voice, data, video, etc. | Smart logistics |
| 14 | Composite Scenario | 0.1-1 Gbps | $10^3$-$10^4$/km | High | 10-100 ms | Control signal, voice, data, video, etc. | Intelligent operation |
| 15 | UAV-to-Ground | 0.01-0.1 Gbps | $10^3$-$10^4$/km$^2$ | High | 10-100 ms | Control signal, voice, data, video, etc. | Intelligent operation and maintenance |
| 16 | Satellite-to-Ground | 0.1-1 Gbps | $10^3$-$10^4$/km$^2$ | High | 100-400 ms | Control signal, voice, data, video, etc. | Intelligent operation |
| 17 | Railroad IoTs | 0.01-0.1 Gbps | $10^4$-$10^5$/km | High | 1-10 ms | Data of sensing and monitoring | Intelligent operation and maintenance |
| 18 | Railway Depot | 0.1-1 Gbps | $10^4$-$10^5$/km$^2$ | High | 1-10 ms | Control signal, voice, data, video, etc. | Intelligent maintenance |

*1. Conventional Scenarios*. Scenarios 1 to 9 in Table 1 can be referred as conventional scenarios, which generally exist in GSM-R. Even though the propagation environments are generally similar to GSM-R, higher carrier frequency and larger bandwidth of 5G-R still result in different channel characteristics compared with the 930 MHz narrowband system of GSM-R. For scenarios 1 to 8, it is highly possible to deploy 5G-R at 2100 MHz to provide control signal transmission, which would lead to more handovers compared with GSM-R at 930 MHz as the coverage of BS is reduced. For scenario 9, mmWave may be used for train data offloading, which has large propagation and diffraction loss, and the scattering environments of train station and platform would result in complex and rich MPCs. Moving speed in the conventional scenarios is generally high and ultra-reliable wireless transmission is required to ensure the safety of train control signal transmission. Another important feature of the conventional scenario is that LOS path generally exists, though in urban and mountain areas, LOS may be occasionally blocked. In the station platform, NLOS paths generally dominate because of the complex structures and scattering environments. Since the typical communication services in most of the conventional scenarios include control signal, voice, data, video, etc., the experienced data rate is generally expected to reach 0.1-1 Gbps with the connection density of $10^3$-$10^4$/km along railway line.

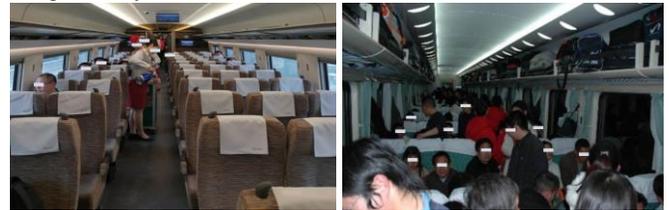

Fig. 1. Example photos of in-carriage scenario. Left – with few passagers. Right – with many passagers.

*2. In-Carriage Scenario*. Scenario 10 has received increasing attention in 5G-R as many communication devices and modules are deployed in this scenario. Fig. 1 shows some example photos of in-carriage scenario. In GSM-R, in-carriage communications are mainly voice services for crew members, which have small user number and data rate. In 5G-R, the demand for in-carriage communication is increasing. In addition to general voice services for crew members and passengers, it should support Internet access services with the connection density of $10^3$-$10^4$/km and up to 0.1-10 Gbps experienced data rate with the latency of 1-10 ms. Internet browsing, video conferencing, and even the wireless high-definition video streaming for on-board entertainment should be supported. The in-carriage scenario involves two communication modes, BS to users inside carriage and access point (AP) to users inside carriage. The former is similar to outdoor-to-indoor scenarios, where propagation link will encounter significant penetration loss, which reduces the received signal strength. For the latter case, the inside-carriage scenario is a closed environment and there are many scatterers such as glass, metal, and seats, which will lead to extremely complex MPC distribution and NLOS propagation. Therefore, since the scattering environments are totally different compared with cellular communications (e.g., the impacts of train car, special reflectors inside carriage, crowded people, huge penetration loss, etc.), the cellular channel models cannot be directly applied to in-carriage scenario. Considering the high penetration loss of carriage, high Doppler shift, frequent handover and other channel features associated with high mobility, high data rate reliable communications of BS to users inside carriage is challenging. One solution is that users inside carriage connect to APs, and APs connect to antennas on the roof of train and further communicate with BSs.

TABLE 2
PROPAGATION CHARACTERISTICS OF EACH SCENARIO

| No. | Scenario | LOS/NLOS | Speed | Typical Frequency Bands | Propagation Characterisitcs |
|---|---|---|---|---|---|
| 1 | Viaduct | LOS | 0-500 km/h | 450 MHz, 900 MHz, 2.1 GHz, 2.6 GHz | Strong LOS |
| 2 | Cutting | LOS | 0-500 km/h | | Rich reflection and scattering components |
| 3 | Tunnel | LOS | 0-500 km/h | | Waveguid effect |
| 4 | Urban | mainly LOS | 0-500 km/h | | Close to cellular channel |
| 5 | Suburban | LOS | 0-500 km/h | | Close to cellular channel |
| 6 | Rural | LOS | 0-500 km/h | | Close to cellular channel |
| 7 | Mountain | LOS/NLOS | 0-500 km/h | | Diffraction |
| 8 | Water | LOS | 0-500 km/h | | Reflection |
| 9 | Platform | LOS/NLOS | 0-80 km/h | 900 MHz, 2.1 GHz, mmWave bands | Closed and semi-closed structure |
| 10 | In-Carriage | LOS/NLOS | 0-10 km/h | 900 MHz, 2.1 GHz, mmWave bands | Closed environment with rich MPCs |
| 11 | Train-to-Train | LOS/NLOS | 0-1000 km/h | Mainly 450 MHz | NLOS dual-mobility with high speed |
| 12 | Station | LOS/NLOS | 0-10 km/h | 900 MHz, 2.1 GHz, mmWave bands | Dynamic pedestrian in channel |
| 13 | Railway Yard | LOS/NLOS | 0-80 km/h | 900 MHz, 2.1 GHz | Rich metal scatterers |
| 14 | Composite Scenario | Mainly LOS | 0-500 km/h | 450 MHz, 900 MHz, 2.1 GHz, 2.6 GHz | Severe fading |
| 15 | UAV-to-Ground | Mainly LOS | 0-500 km/h | 800 MHz, 1.4 GHz, 2.4 GHz, 5.8 GHz, mmWave bands | Strong LOS |
| 16 | Satellite-to-Ground | Mainly LOS | 0-500 km/h | 1-2 GHz, 4-8 GHz, 12-18 GHz | Strong LOS |
| 17 | Railroad IoTs | LOS/NLOS | - | 400-900 MHz, 1.8 GHz, 2.1 GHz, 2.4 GHz | Near-ground propagation |
| 18 | Railway Depot | LOS/NLOS | - | 2.1 GHz, mmWave bands | Rich metal scatterers |

*3. Train-to-Train Scenario*. Train-to-train (T2T) communication enables information interaction between trains, including speed, location and other safety-related information, so as to avoid transportation accidents and improve efficiency. The communication distance of T2T is generally over tens of kilometers. It involves diversified and composite scenarios (e.g., viaduct, suburban, tunnel, etc.) and long communication distance, which result in significant challenges in channel prediction and communication delay. Another important feature of T2T communication is NLOS dual-mobility with high speed, which leads to extremely time-varying and non-stationary channel and reduces communication quality. Nevertheless, it requires high reliability and low latency (1-10 ms) communications due to safety concerns for control signal of intelligent operation. In order to ensure reliable T2T communications in composite railway environments, it is important to select appropriate frequency band. Since T2T communication often occurs under NLOS condition, the carrier frequency of the system should have good diffraction property [18].

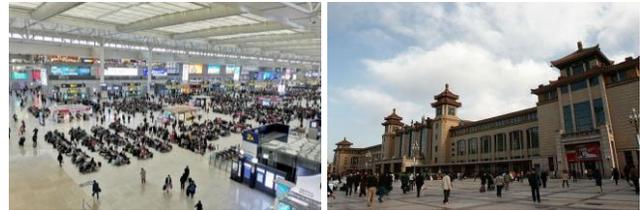

Fig. 2. Example photos of railway stations. Left – the closed area of station hall. Right – the open area of station square.

*4. Station Scenario*. Railway station includes many users and devices which should be supported by 5G-R systems. On one hand, the smart station requires a large number of devices (around $10^4$-$10^5$/km$^2$ connection density) to be deployed so as to improve safety, environmental comfort, and operational efficiency. On the other hand, massive passengers in station require a well-designed system to provide wireless services with fairly high data rate (up to 10 Gbps) and a large number of access devices. Station scenario generally includes closed area and open area as shown in Fig. 2, and consists of both LOS and NLOS propagation. The closed area mainly includes station hall, which has densely distributed people and frequent pedestrian movements. Due to the large number of users, high traffic is expected in this environment. The closed structure of station hall generally increases signal strength, results in rich MPCs, and leads to significant challenges of BS deployments and interference management. The open area of station mainly includes railway station square, which has dense people and open space. There are many pedestrian in this area leading to extra scattering components. The requirement of wireless service in railway station square is similar to station hall, and the open structure in this area provides more solutions for BS deployments.

*5. Railway Yard Scenario*. Railway yard is an important scenario in railway traffic, which generally includes freight yard, marshaling yard, shunting yard, etc, as shown in Fig. 3. This scenario usually exists at some railway freight stations and is used for the formation of freight trains, and/or with a large volume of traffic, and/or with extensive track systems. In railway yard, freight trains that consist of isolated cars are made into trains and divided according to different destinations. In 5G-R, many sensors and access devices are deployed in railway yard scenario with a connection density of $10^4$-$10^5$/km$^2$, such as on railway cars, tracks, roadside infrastructures, freights, etc. Complete wireless coverage needs to be provided to improve efficiencies and safeties of loading/unloading, railway car separation and sorting, and yard switcher operation, so as to further support smart traffic and smart logistics applications with high reliability and low latency (around 1-10 ms). In railway yard scenario, huge traffic of controlling signal is expected and a large number of access devices should be supported. In addition, a number of metallic objects such as train car and freight container result in rich reflection and scattering components.

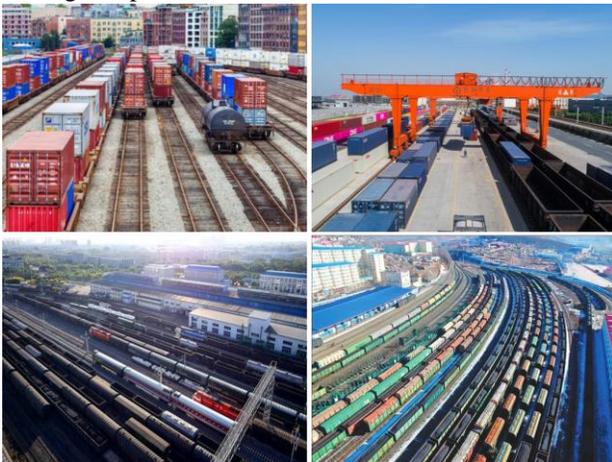

Fig. 3. Example photos of railway yard scenario.

*6. Composite Scenario*. In 5G-R, cell size is reduced because of using higher carrier frequency. Since different complicated environments exist along railway lines, the propagation channel for one cell cannot be characterized by one single scenario but a composite one with multiple scenarios. The common composite scenarios include the combination of bridges, tunnels, cuttings, and station, within the range of one cell. The most typical composite scenario in 5G-R is tunnel group scenario. It mostly appears when train is moving in mountain environment, where multiple short tunnels exist in railway line and train frequently moves out of a tunnel and enters another one. Such frequent changes of propagation environments lead to significant fading and reduced transmission quality. The tunnel group scenario should have LOS propagation by well-designed coverage planning so as to guarantee good communication quality and wireless link may frequently switch between APs and infrastructures such as BS and relay.

*7. UAV-to-Ground Scenario*. For 5G-R, it is possible to deploy UAVs across a number of application domains, ranging from wireless transmission improvement, railway environmental monitoring, coverage enhancement, and remote sensing, etc. Furthermore, UAVs can also provide wireless local area network services to railway users. The flight speed of UAV ranges from 10 m/s to 100 m/s, and UAV can act as aerial BS for railway coverage enhancement in some complex environments. It has the advantage of having high LOS probability by increasing UAV height and good transmission quality can be guaranteed (0.1-1 Gbps with high reliability connections). However, airframe shadowing may lead to extra attenuation, which should be considered in link budget. For UAV-to-ground scenario, MPCs are caused by earth-face reflection and scattering from ground objects, which are affected by complex railway environments. However, radio signal generally experiences less scattering because of the absence of surrounding objects at UAV side. Therefore, the number of MPCs tends to be less for UAV channels compared with land mobile communications.

*8. Satellite-to-Ground Scenario*. Satellite-to-ground communication system will be integrated with railway communication network in 5G-R era, and further establish space-air-ground integrated networks. Satellite can support and improve various applications for 5G-R such as train precise localization, railway emergency communications, user broadband internet, and backhaul for terrestrial BS/AP, etc. Moreover, it can provide wireless transmission services (e.g., voice, data, and video) with 0.1-1 Gbps data rate, 100-400 ms latency, and high reliability. The propagation channel of satellite-to-ground communication is affected by several factors such as the surrounding environment, atmospheric attenuation, rain attenuation, fog attenuation, etc., especially for high frequency bands. In integrated networks, long-distance satellite link transmission may experience different propagation media, such as stratosphere, troposphere and outer space, and the different propagation characteristics will make the characterization of satellite-ground channel more complicated.

*9. Railroad Internet-of-Things Scenario*. In 5G-R, trains are designed to travel with a speed up to 500 km/h, thus the safety is the top priority issue. The monitoring system based on railroad Internet-of-Things (IoTs) network is important to establish a warning system and avoid accidents such as landslide and railway foreign object intrusion. The IoT based monitoring system (including numerous sensors around $10^4$-$10^5$/km$^2$) usually transmits monitoring data through cellular network such as 5G-R or through railway optical fiber system. Benefit by fiber channel, the optical fiber based railway monitoring system shows high robustness on transmission performance (0.01-0.1 Gbps data rate with 1-10 ms latency), however, deploying optical fiber system requires high cost. On the other hand, integrating railway IoT into 5G-R has more flexibility for implementation. The sensors of railway IoT are usually deployed alone with railway line, and the propagation channels are affected by the complex railway environments. Hence, a well-designed railroad IoT system is critical to safety.

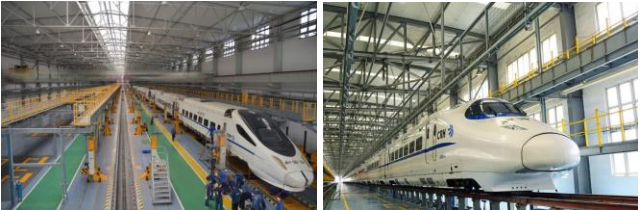

Fig. 4. Example photos of railway depot scenario.

*10. Railway Depot Scenario.* This scenario is the place where locomotives are serviced and maintained. In 5G, smart manufacturing is developed to realize the potential of 5G network with Industry 4.0 to enable intelligent automation and leverage real time data across operations. 5G-R networks offer to build smart factories in railway depot scenario and truly take advantage of technologies such as automation, artificial intelligence, augmented reality for troubleshooting, and IoT. In railway depot scenario, low latency of 1-10 ms and high reliability are needed to support critical railway manufacturing applications. High bandwidth and connection density are required to secure ubiquitous connectivity. 5G-R technology in railway depot will allow for higher flexibility, lower cost, and shorter lead times for smart factory maintenance, reconfiguration, layout changes, and alterations. In railway depot scenario, huge traffic of mechanical controlling signals and large access devices (around $10^4$-$10^5$/km$^2$) are expected. The propagation environment is similar to warehouse and factory workshop and results in complex and rich reflection and scattering components.

## IV. CONCLUSIONS

In this paper, a 5G-R oriented standardized radio communication scenario classification including 18 scenarios is proposed. The proposed classification is based on the physical environment, railway service requirements, and radio propagation characteristics. This paper points out that the scenarios in 5G-R include the conventional scenarios such as urban, tunnels and viaducts, and some special scenarios which have huge requirements of smart services and are important for the evolution of railway communication systems, such as UAV-to-Ground, train-to-train, railroad IoTs, and railway yard. This paper presents some key attributes of communication scenarios, such as LOS/NLOS, frequency bands, and propagation characteristics. For the evolution and deployment of 5G-R networks, the proposed standardized scenario classification can play an essential role in promoting channel modeling and system performance evaluation.

## ACKNOWLEDGMENT

This work is supported by National Key R&D Program of China under Grant 2020YFB1806903, National Natural Science Foundation of China under Grant 61922012, 61771037, 62001519, and 61961130391, and State Key Laboratory of Rail Traffic Control and Safety under Grant RCS2020ZT008.

7